# Investigation of structural and optoelectronic properties of $BaThO_3$

*G. Murtaza[1], Iftikhar Ahmad[1,] *, B. Amin[1], J. Maqssod[1], A. Afaq[2], M. Maqbool[3], I. Khan[1], M. Zahid[1]*

1. *Materials Modeling lab, Department of Physics, Hazara University, Mansehra, Pakistan*
2. *Center for Solid State Physics, University of the Punjab, Lahore, Pakistan*
3. *Department of Physics and Astronomy, Ball State University, Muncine, Indiana, USA*

## ABSTRACT

Structural and optoelectronic properties of $BaThO_3$ cubic perovskite are calculated using all electrons full potential linearized augmented plane wave (FP-LAPW) method. Wide and direct band gap, 5.7 eV, of the compound predicts that it can be effectively used in UV based optoelectronic devices. Different characteristic peaks in the wide UV range emerges mainly due to the transition of electrons between valance band state O-p and conduction band states Ba-d, Ba-f, Th-f and Th-d.

* *Corresponding author's email: ahma5532@gmail.com*

## 1. Introduction

Knowledge of physical properties of materials for their possible applications has always been a prime field of interest. Materials scientists even in today's technologically advanced society are still struggling for chief and efficient materials. Perovskites family contains a large number of compounds ranges from insulators to superconductors and from diamagnetic to colossal magneto-resistive (CMR) compounds. It is also expected that the ideal materials for spintronics applications could be double perovskites.

Perovskites make huge portion of the mantle of the earth crust and therefore the investigation of physical properties of these compounds is highly desirable. They reveal many fascinating properties from both theoretical as well as experimental point of view. High thermoelectric power, ferroelectricity, superconductivity, charge ordering, spin dependent transport, colossal magneto-resistance, and the interplay of structural, magnetic and optical properties are commonly observed features in these materials. These materials are frequently used as sensors, substrates, catalytic electrodes in fuel cells and are also promising candidates for optoelectronics [1-9].

Barium thorate, $BaThO_3$, is one of the products of fission reactor and due to its importance in nuclear reactor Moreira et al. [10] studied its structural properties and Mishra et al. [11] studied its thermodynamic stability. Other physical properties of $BaThO_3$ like heat capacity [12] and Gibbs free energy [13,14] are also reported. Though barium thorate is an important member of the Perovskite family but even then no theoretical work has been published yet.

In the present article, structural and optoelectronic properties of $BaThO_3$ cubic perovskite have been investigated. Lattice constant, bulk modulus, derivative of bulk modulus, band gap, density of states, complex dielectric function, complex refractive index, reflectivity, optical conductivity, energy loss function and absorption coefficient of the compound are calculated for the first time using self consistent full potential linearized augmented plane wave (FPLA-PW) method with Wu-Cohen generalized gradient approximation (GGA). It is expected that the present work will help in the understanding of optoelectronic behavior of $BaThO_3$ and the article will also cover the lack of theoretical data on the structural, electronic and optical properties of this interesting compound.

## 2. Theory and computation

Structural parameters of $BaThO_3$ are evaluated by fitting the unit cell energy versus unit cell volume using Birch-Murnaghan's equation of state [15]:

$$E(V) = E_0 + \frac{9V_0B_0}{16}\left[\left\{\left(\frac{V_0}{V}\right)^{\frac{2}{3}} - 1\right\}^3 B_0^{/} + \left\{\left(\frac{V_0}{V}\right)^{\frac{2}{3}} - 1\right\}^2 \left\{6 - 4\left(\frac{V_0}{V}\right)^{\frac{2}{3}}\right\}\right] \quad (1)$$

where $E_0$ is the total energy of the unit cell, $V_0$ is the unit cell volume, $B_0$and $B_0^{/}$ are the bulk modulus at zero pressure and its derivative with pressure. Real and imaginary parts of dielectric function, $\varepsilon_1(\omega)$ and $\varepsilon_2(\omega)$, are calculated by the following relations [16, 17]:

$$\varepsilon_2(\omega) = \frac{8}{2\pi\omega^2}\sum_{n\acute{n}}\int |P_{n\acute{n}}(k)|^2 \frac{dS_k}{\nabla\omega_{n\acute{n}}(k)} \quad (2)$$

$$\varepsilon_1(\omega) = 1 + \frac{2}{\pi}P\int_0^\infty \frac{\acute{\omega}\,\varepsilon_2(\acute{\omega})}{\acute{\omega}^2 - \omega^2}d\acute{\omega} \quad (3)$$

where $P_{n\acute{n}}(k)$ is the dipole matrix element between initial and final states, $S_k$ is an energy surface with constant value, $\omega_{n\acute{n}}(k)$ is energy difference between two states and p denotes the principal part of the integral. Real and imaginary parts of dielectric function are used to calculate refractive index and extension coefficient using relations:

$$n(\omega) = \frac{1}{\sqrt{2}}\left[\{\varepsilon_1(\omega)^2 \ + \varepsilon_2(\omega)^2\}^{1/2} + \varepsilon_1(\omega)\right]^{1/2} \tag{4}$$

$$k(\omega) = \frac{1}{\sqrt{2}}\left[\{\varepsilon_1(\omega)^2 \ + \varepsilon_2(\omega)^2\}^{1/2} - \varepsilon_1(\omega)\right]^{1/2} \tag{5}$$

Real and imaginary parts of refractive index are further used in the following relation to evaluate normal incident reflectivity of $BaThO_3$:

$$R(\omega) = \left|\frac{\tilde{n}-1}{\tilde{n}+1}\right| = \frac{(n-1)^2 + k^2}{(n+1)^2 + k^2} \tag{6}$$

Similarly energy loss function $L(\omega)$[18], absorption coefficient $\alpha(\omega)$ and frequency dependent optical conductivity $\sigma(\omega)$ are calculated by the following relations:

$$L(\omega) = Im\left(-\frac{1}{\varepsilon(\omega)}\right) \tag{7}$$

$$\alpha(\omega) = \frac{4\pi k(\omega)}{\lambda} \tag{8}$$

$$\sigma(\omega) = {2W_{ev}\hbar\omega}/{\vec{E}_0} \tag{9}$$

where $W_{ev}$ is transition probability per unit time.

In the present density functional calculations full potential linearized augmented plane wave (FP-LAPW) method with Wu-Cohen generalized gradient approximation [19], implemented in the wien2k code [20], is used to solve Kohan-Sham equations [21] for $BaThO_3$. In the full potential scheme wave function, potential and charge density are expanded into two different basis. The wave function is expanded in spherical harmonics in the atomic spheres while outside the spheres (interstitial region) it is expanded in plane wave basis. The potential is also expanded in the same manner:

$$V(r) = \begin{cases} \sum_{lm} V_{lm}(r) Y_{lm}(r) \ldots\ldots\ldots\ldots\ldots\ldots\ldots\ldots\ldots\ldots\ldots\ldots\ldots (a) \\ \sum_{k} V_k e^{ikr} \ldots\ldots\ldots\ldots\ldots\ldots\ldots\ldots\ldots\ldots\ldots\ldots\ldots\ldots\ldots (b) \end{cases} \quad (10)$$

where Eq. 10(a) is for inside and 10(b) is for outside of the atomic sphere. Inside the sphere the maximal value of $l$ for the wave function expansion is $l_{max}$ = 10 and is spherically symmetric while outside the sphere it is constant. $R_{MT}$ is chosen in such a way that there is no charge leakage from the core and the total energy convergence is ensured. $R_{MT}$ values of 2.5, 2.19, 1.94 a.u. are used for Ba, Th, O respectively. For wave function in the interstitial region the plane wave cut-off value of $K_{max}$ = 7/ $R_{MT}$ is chosen. 2000 k points are used in the Brillouin zone integration and convergence is checked through self consistency. The perovskite structure of $BaThO_3$ is shown in Fig.1. Ba atom in the unit cell is at (0,0,0), Th atom at (0.5,0.5,0.5) while O atoms at the faces of the unit cell, (0.5,0.5,0), (0.5,0,0.5), (0,0.5,0.5).

## 3. Results and Discussion

Structural and optical properties of $BaThO_3$ are calculated by all electrons ab-initio full potential linearized augmented plane wave (FP-LAPW) method. The unit cell volume of $BaThO_3$ is varied in the neighborhood of the experimental value 4.48 Å [10] and the corresponding energies are calculated using Wu-Cohen GGA scheme. Birch-Murnaghan's equation of state [15] is used to plot, relation between energy and volume shown in the Fig. 2. From the plot relaxed lattice constant is obtained by taking the volume corresponding to the minimum energy. The optimized parameters are quoted in Table1and are compared with the other available data. It is evident from the table that our calculated lattice parameter is in good agreement to the experimental [10] and other theoretical results [31, 32]. The large value (124 GPa) of bulk modulus suggest that $BaThO_3$ is less compressible with pressure derivative of 2.41.

Band structure and density of states are calculated with the well converged self consistent solution of FP-LAPW, shown in Figs. 3 and 4. It is clear from Fig. 3 that, $BaThO_3$ is a direct band gap compound with a band gap of 5.7 eV at the Γ symmetry point. The calculated direct band gaps at other symmetry points are also presented in Table 1. Wide and direct band gap at the Γ symmetry point has also been reported for similar compounds like $BaTiO_3$ [22], $BaZrO_3$ [23] and $BaHfO_3$ [24] which confirms the correctness of our band structure calculation for $BaThO_3$.

The calculated density of states for $BaThO_3$ is shown in Fig. 4. It is clear from the figure that the density of states can be divided into three regions. The first region is from -13.57 eV to -11.48 eV with a major contribution from Ba-5p states while small contribution from Th-6p states. The next region is valence band, ranges from -3.7 eV to 0.0 eV and the main contribution is due to O-2p states. The last region is conduction band, which ranges from 3.33 eV to 13.66 eV. In this band, the prominent contribution comes from Ba-4f states with a noticeable part from Ba-5d, Th-5f and Th-6d states.

In order to envision the bonding nature and charge transfer the charge density in the (100) and (110) planes is plotted in Fig. 5(a, b). From Fig. 5(a), it can be seen that there is a large transfer of charge among Ba and O atoms with a very small contour extension among the ions and hence Ba–O bond is strongly ionic with very weak covalent nature. Fig. 5(b) shows that electrons are strongly shared and distributed along the Th–O bond. Hence, their bond is strongly covalent with a bond length 2.27 A˚. Similar charge distribution and bonding nature has been reported for a similar compound, $BaHfO_3$ [24] .

Optical spectra of a compound provide useful information about the internal structure of that material. Keeping in view the importance of optical properties of $BaThO_3$, different frequency dependent optical parameters are described in terms of electronic properties to recognize the optoelectronic nature of the compound for different devices applications.

The dielectric function,$\varepsilon(\omega) = \varepsilon_1(\omega) + i\varepsilon_2(\omega)$, describes the optical response of the medium to the incident photons, $E = \hbar\omega$. The imaginary part of the dielectric function, $\varepsilon_2(\omega)$, is the most important parameter of the optical properties of a material and therefore its details study is important. The variation of the imaginary part of the dielectric function with the incident photon energy is shown in Fig. 6. It is clear from the figure that the critical point of $\varepsilon_2(\omega)$ is around 5.3 eV. Two high peaks A and C at 8.7 eV and 14.9 eV along with some noticeable peaks B, D, E, F and G are located at 12.5, 17.9, 18.5, 20 and 26 eV respectively. The origin of these peaks lies in the inter band transitions which can be related to the density of states of the compound shown in Fig. 4. It can be seen in the figure that the peaks A, B, C and D are due to the transition of electrons from O-p states to Ba-d, Th-f, Ba-f and Th-d states respectively. While the peaks E, F and G emerges due to the transition of electrons from mixed Ba-p and Th-p states below the valence band to Th-F, Ba-f and Th-d states in conduction band respectively.

The optical band gap obtained from the imaginary part of the dielectric function is 5.6 eV which is close to the band gap obtained from the band structure (5.7 eV). It is well understood that materials with band gaps larger than 3.1 eV works well in the ultraviolet (UV) region of the spectrum [25-29]. Hence this direct and wide band gap material could be suitable for the high frequency UV devices applications. As the compound also shows strong absorbing nature in the different parts of spectra, ranges 7- 27 eV. So, it can also be used as a filter for various energies in the UV spectrum.

Different important frequency dependent optical parameters like; real part of dielectric function $\varepsilon_1(\omega)$, extinction coefficient $k(\omega)$, absorption coefficient $\alpha(\omega)$, energy loss function $L(\omega)$, refractive index $n(\omega)$, reflectivity $R(\omega)$ and optical conductivity $\sigma(\omega)$are plotted in Fig. 7. The plot of the real part of the dielectric function (Fig.7a) explains different important physical aspects of $BaThO_3$. It is clear from the figure that zero frequency limit, $\varepsilon_1(0)$, is 2.95. The magnitude of $\varepsilon_1(\omega)$ increases from zero frequency limit to peak value 7.9 at 6.9 eV. After maximum value it starts decreasing and becomes flatten with small variations. It is further noted that the real part of dielectric function becomes negative in the energy ranges 15.19−15. 6 3 eV, 16.01− 16.39 eV, 18.65− 19.06 eV, 20.01− 23.68 eV and 26.62–28.15 eV. As materials behave metallic for negative values of $\varepsilon_1(\omega)$ and are dielectric otherwise [30]. Therefore for these ranges of energy $BaThO_3$ is metallic. Interestingly, the refractive index (Fig. 7e) follows the trend of $\varepsilon_1(\omega)$. The zero frequency limit of the refractive index is 1.7 with a maximum value of 2.7 at 8.1 eV.

The figure also reveals similarities in the trends of extinction coefficient $k(\omega)$ and absorption coefficient $\alpha(\omega)$ (Fig. 7b and 7c). The Gaussian smearing value in the present calculations is 0.1 eV. The critical point of $k(\omega)$ is 6.3 eV and of $\alpha(\omega)$ is 6.8 eV. Both have strong response to the incident photons in the range 8-29 eV. The highest peak value of extinction coefficient is 8.7 and of absorption coefficient is $218 \times 10^{-4} cm^{-1}$ at 20.2 eV.

Frequency corresponding to the plasma resonance can be calculated from the energy loss spectra, shown in Fig. 7(d). The maximum resonant energy loss is at 28.5 eV which corresponds to plasma frequency $2.7 \times 10^{15}$Hz. From Fig. 7(f), the zero frequency limit of reflectivity for $BaThO_3$ is found to be 0.065. There are high reflection peaks at energies 8.7 eV, 15.25 eV, 22.1 eV and 27.8 eV corresponding to the negative values of $\varepsilon_1(\omega)$ . The optical conductivity (Fig.7g) starts responding to the applied energy field from 6.5 eV. The good response is found in the range 7 to 28eV. Maximum optical conductivity of the compound is at 14.87 eV of magnitude $9288\Omega^{-1}\ cm^{-1}$.

## 4. Conclusion

Density functional FP-LAPW calculations are performed on $BaThO_3$ for the first time to investigate structural and optoelectronic properties. The calculated lattice constant 4.55 Å is in good agreement with the available data. The calculated band structure predicts that the compound has a wide and direct band gap of 5.7 eV. On the basis of the direct wide band gap and the spectra of imaginary part of the dielectric function it can be concluded that this material will be useful for optoelectronic devices in the UV region of the spectrum.

**Acknowledgements**

Prof. Dr. Keith Prisbrey, MSE, University of Idaho and Prof. Dr. Nazma Ikram, Ex. Director, Center for Solid State Physics, Punjab University are highly acknowledged for their valuable suggestions.

## References

[1] S. Moskvin, A. A. Makhnev, L. V. Nomerovannaya, N. N. Loshkareva, and A. M. Balbashov, Phys. Rev. B 82, 035106 (2010)

[2] C. Weeks and M. Franz, Phys. Rev. B 82, 085310 (2010).

[3] M. L. Scullin, C. Yu, M. Huijben, S. Mukerjee, J. Seidel, Q. Zhan, J. Moore, A. Majumdar, R. Ramesh, Appl. Phys. Lett. 92 (2008) 202113.

[4] X. Y. Zhou, J. Miao, J. Y. Dai, H. L. W. Chan, C. L. Choy, Y. Wang, Q. Li, Appl. Phys. Lett. 90 (2007) 012902.

[5] J. Lettieri, M. A. Zurbuchen, Y. Jia,a) and D. G. Schlom, S. K. Streiffer, M. E. Hawley, Appl. Phys. Lett.76 (2000) 2937.

[6] M. Woerner ,C. K. Schmising, M. Bargheer, N. Zhavoronkov, I. Vrejoiu, D. Hesse, M. Alexe, T. Elsaesser, Appl. Phys. A 96 (2009) 83.

[7] S. Ghosh, S. Dasgupta, Materials Science-Poland. 28 (2010) 427.

[8] T. Yamada, C. S. Sandu, M. Gureev, V. O.Sherman, A. Noeth, P. Muralt, A. K. Tagantsev, N. Setter, Adv. Mater. 21 (2009) 1363.

[9] J. M. D. Coey, M. Viret, S. V. Molnaacuter, Advances in Physics. 48 (1999) 167.

[10] R. L. Moreira and A. Dias, J. Phys. Chem. Solids. 68 (2007) 1617.

[11] R. Mishra, M. Ali (Basu), S.R. Bharadwaj, A.S. Kerkar, D. Das, S.R. Dharwadkar, Journal of Alloys and Compounds. 290 (1999) 97.

[12] R. V. Krishnan, K. Nagarajan, P. R. V. Rao, J. Nucl. Mater.99 (2001) 28.

[13] S.R. Bharadwaj, R. Mishra, M. Ali (Basu), D. Das, A.S. Kerkar, S.R.Dharwadkar, J. Nucl. Mater. 275 (1999) 2001.

[14] R. D. Purohit, A. K. Tyagi, M. D. Mathews, S. Saha, J. Nucl. Mater. 280 (2000) 51.

[15] Francis Birch, Physical Review. 71 (1947) 809.

[16] M. A. Khan, A. Kashyap, A. K. Solanki, T. Nautiyal, S. Auluck. Phy. Rev. B 23 (1993) 1697.

[17] F. Wooten, Optical properties of Solids, Academic Press, New York, 1972.

[18] D. Groh, R. Pandey, M. B. Sahariah, E. Amzallag ,I. Baraille, M. Reart, Journal of Physics and Chemistry of Solids.70 (2009) 789.

[19] Z. Wu, R. E. Cohen, Phys. Rev. B 73 (2006) 235116.

[20] P. Blaha, K. Schwarz, G.K.H. Madsen, D. Kvasnicka, J. Luitz, WIEN2k, An Augmented Plane Wave Plus Local Orbitals Program for Calculating Crystal Properties, Vienna University of Technology, Vienna, Austria, 2001., K. Schwarz, P. Blaha, G.K.H. Madsen, Comp. Phys. Commun. 147 (2002) 71.

[21] W. Kohn, L. J. Sham, Phys. Rev. 140 (1965) A1133.

[22] C. B. Samantaray, H. Sim and H. Hwang, Microelectronics Journal, 36 (2005) 725.

[23] R. Khenata, M. Sahnoun, H. Baltache, M. Rerat, A. H. Rashek, N. Illes and B. Bouhaf. Solid State Communications, 136 (2005) 120.

[24] H. Zhao, A. Chang and Y. Wang, Physica B: Condensed Matter 404 (2009) 2192.

[25] M. Maqbool, I. Ahmad, H. H. Richardson and M. E. Kordesch, Appl. Phys. Lett. 91 (2007) 193511.

[26] M. Maqbool, B. Amin and I. Ahmad, J. Opt. Soc. Am. B 26 (2009) 2180.

[27] Muhammad Maqbool, Martin. E. Kordesch, and A. Kayani, J. Opt. Soc. Am. B 26 (2009) 998.

[28] M. Maqbool and I. Ahmad, Current Applied Physics. 9 (2009) 234.

[29] B. Amin, I. Ahmad, M. Maqbool, Journal of Lightwave Technology, 28 (2010) 223.

[30] B. Xu, X. Li, J. Sun, L. Yi. Eur. Phys. J. B 66 (2008) 483.

[31] L. Q. Jiang, J. K. Guo, H. B. Liu, M. Zhu, X. Zhou, P. Wu, C. H. Li, J. Phys. Chem. Solids. 67 (2006) 1531.

[32] A. S. Verma, A. Kumar, and S. R. Bhardwaj, Phys. Stat. Sol. (b) 245 (2008) 1520.

## Figure Captions

**Fig.1**: Crystal structure of cubic perovskite $BaThO_3$.

**Fig.2**: Variation of total energy as a function of unit cell volume of $BaThO_3$

**Fig.3**: Band structure of $BaThO_3$.

**Fig.4**: Total density of states of $BaThO_3$.

**Fig.5**: Total electron density for $BaThO_3$: (a) in the (100) plane (b) in the (110) plane

**Fig.6**: Imaginary part of dielectric function as a function of energy for $BaThO_3$.

**Fig.7**: Frequency dependent optical parameters for $BaThO_3$.

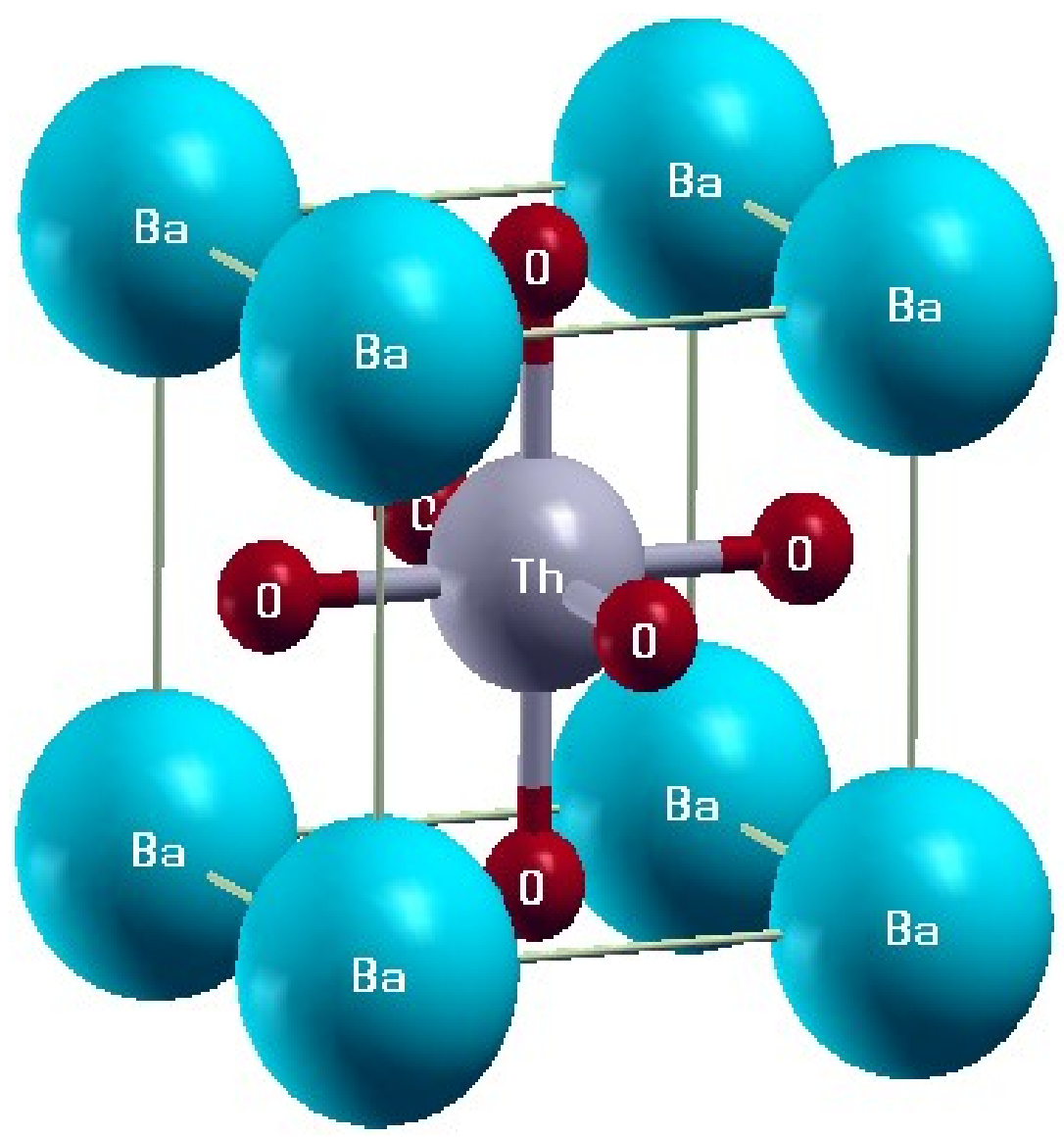

**Figure 1**

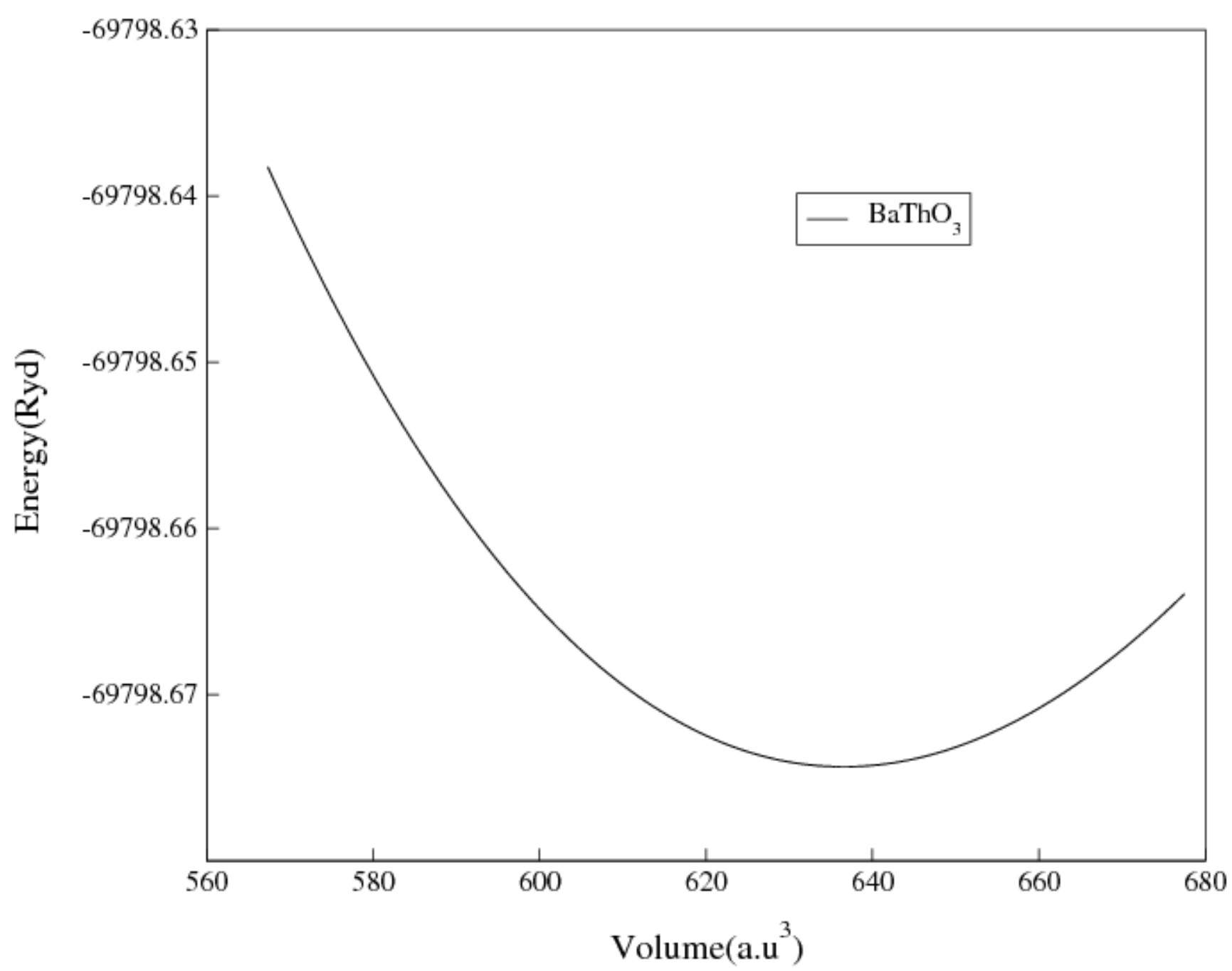

-69798.63
-69798.64
-69798.65
-69798.66
-69798.67
Energy(Ryd)
BaThO3
560
580
600
620
640
660
680
Volume(a.u3)

**Figure 2**

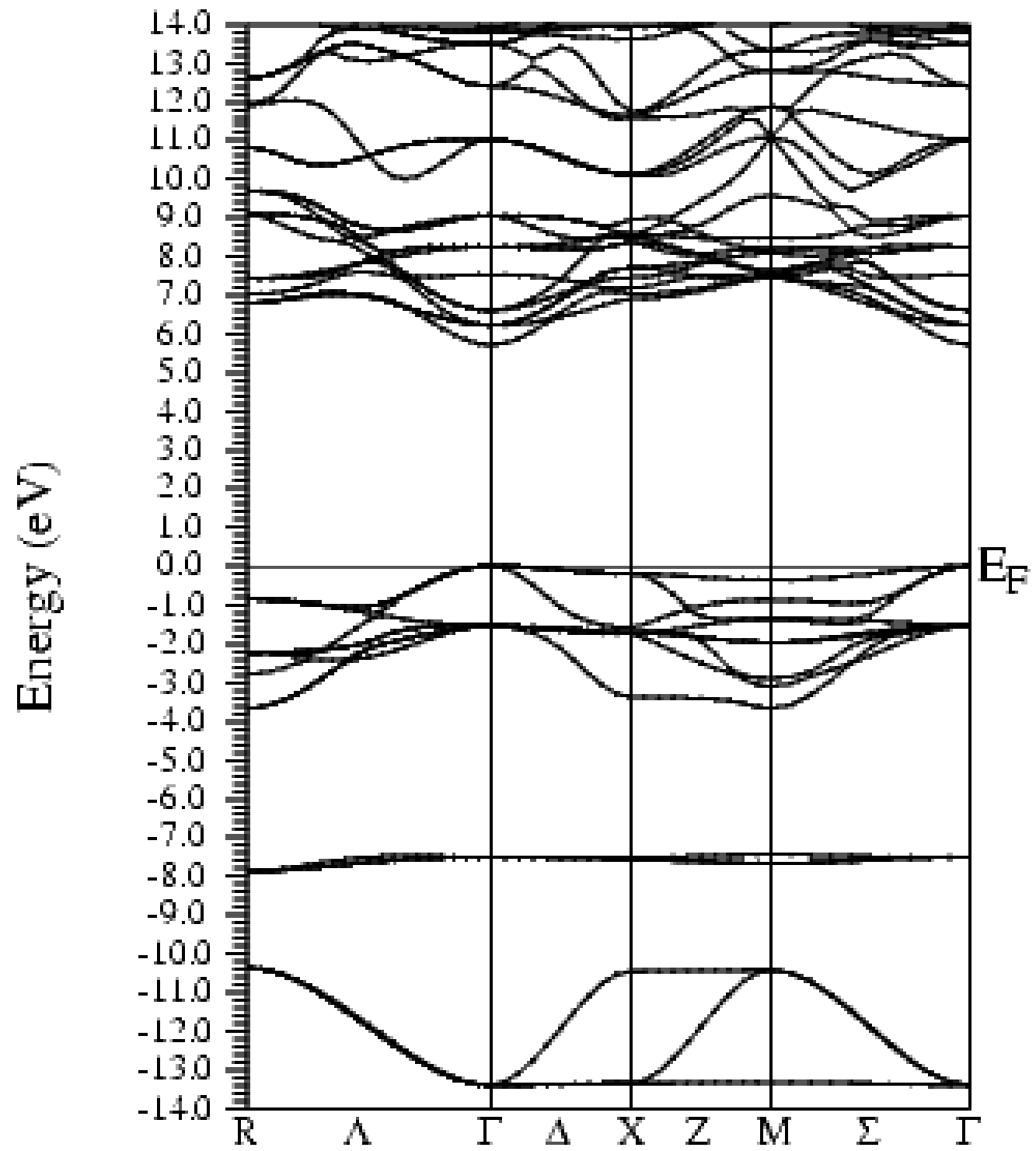

Energy (eV)
14.0
13.0
12.0
11.0
10.0
9.0
8.0
7.0
6.0
5.0
4.0
3.0
2.0
1.0
0.0
-1.0
-2.0
-3.0
-4.0
-5.0
-6.0
-7.0
-8.0
-9.0
-10.0
-11.0
-12.0
-13.0
-14.0
$E_F$
R Λ Γ Δ X Z M Σ Γ

**Figure 3**

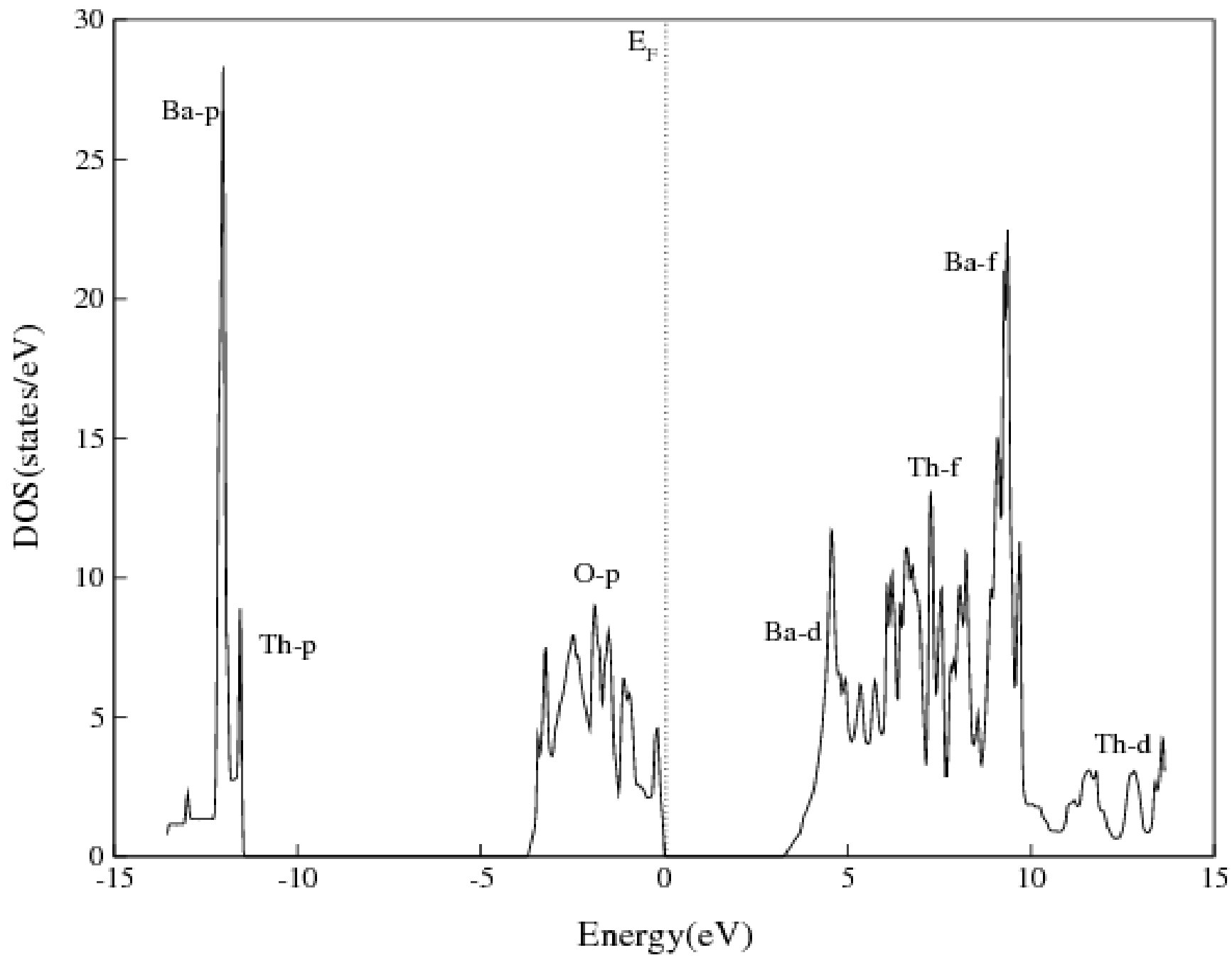

30
25
20
15
10
5
0
DOS(states/eV)
-15
-10
-5
0
5
10
15
Energy(eV)
$E_F$
Ba-p
Th-p
O-p
Ba-d
Th-f
Ba-f
Th-d

**Fig.4**

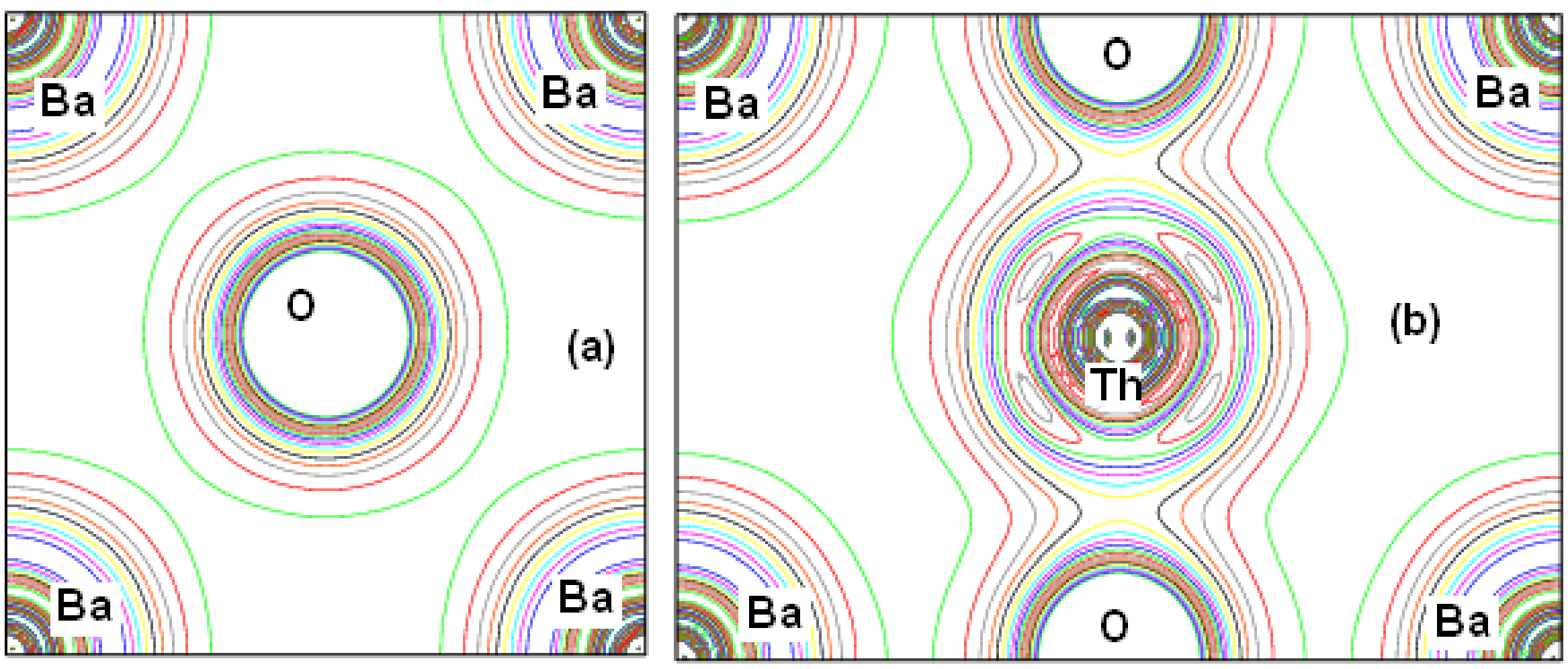

**Figure 5**

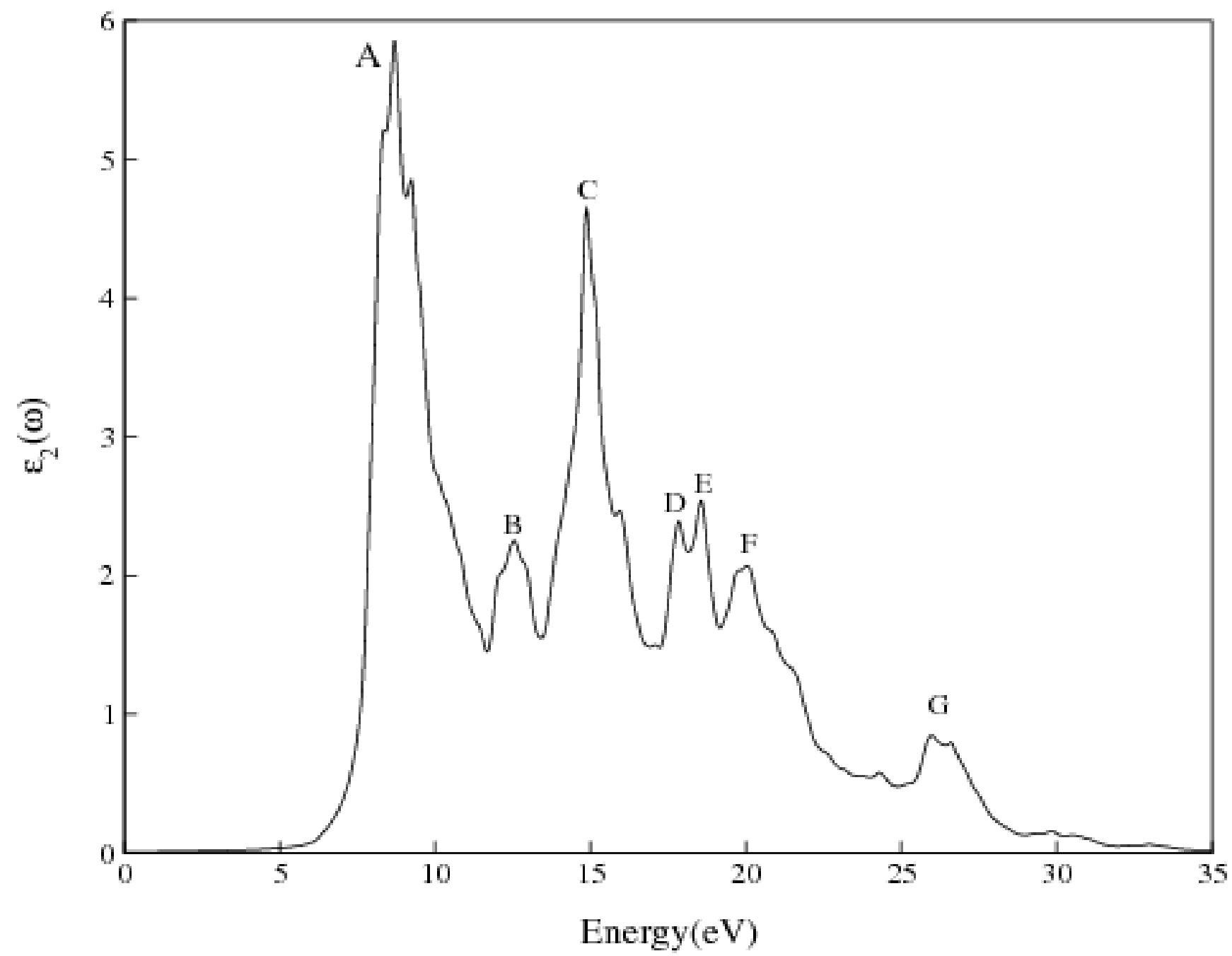
A
B
C
D
E
F
G
$\varepsilon_2(\omega)$
Energy(eV)
0
1
2
3
4
5
6
0
5
10
15
20
25
30
35

**Fig.6**

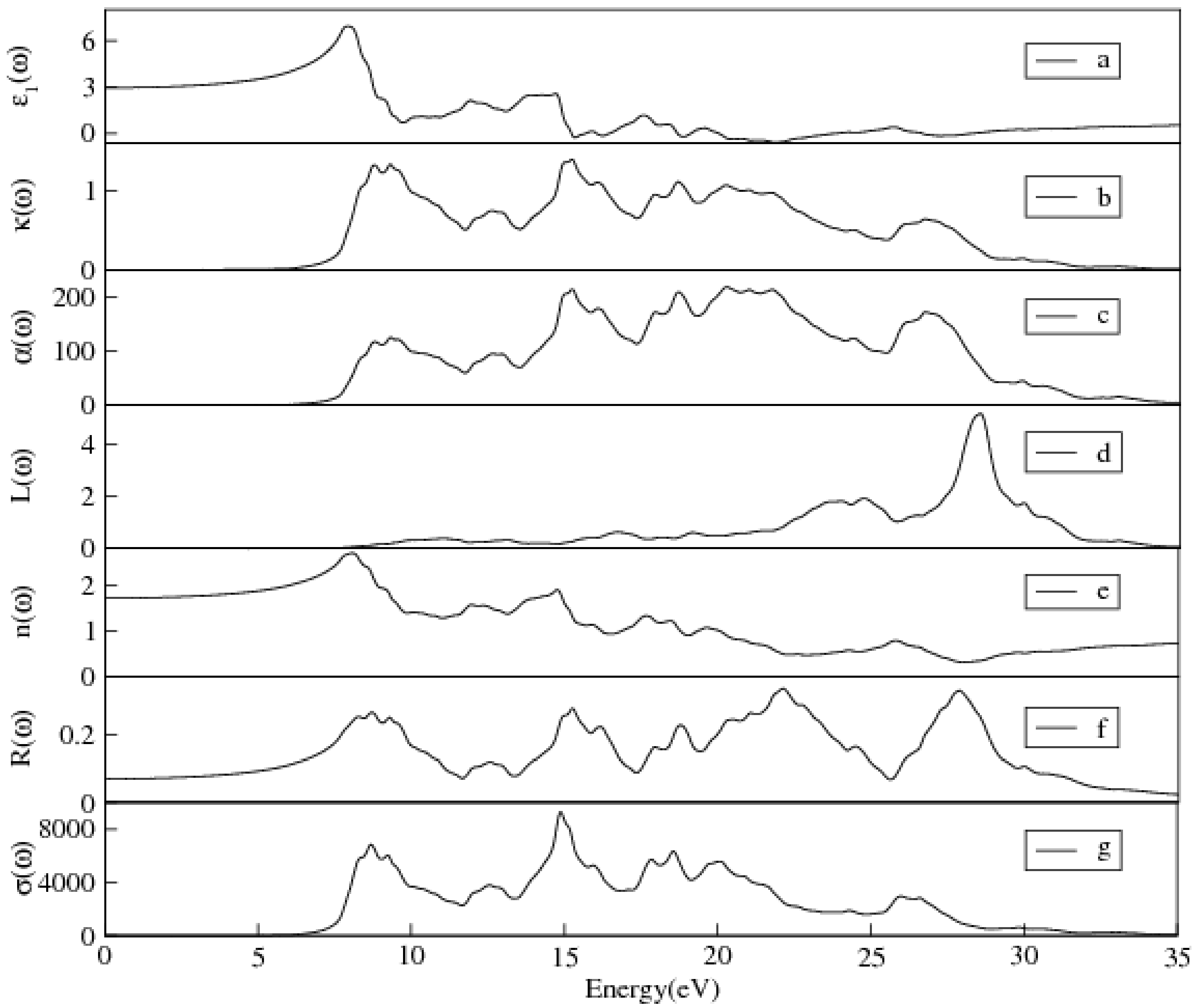

ε1(ω)
6
3
0
a
κ(ω)
1
0
b
α(ω)
200
100
0
c
L(ω)
4
2
0
d
n(ω)
2
1
0
e
R(ω)
0.2
0
f
σ(ω)
8000
4000
0
g
0
5
10
15
20
25
30
35
Energy(eV)

**Fig.7**

**Table 1:** Lattice constant, bulk modulus, pressure derivative of bulk modulus, Band gap of $BaThO_3$ .

| | **This Work** | **Experimental work** | **Other work** |
|---|---|---|---|
| Lattice constant $\boldsymbol{a}(\dot{\mathbf{A}})$ | 4.55 | 4.48[I] | 4.43[II], 4.53[III] |
| Bulk modulus B(GPa) | 124.34 | | |
| Derivative of bulk modulus $\boldsymbol{\acute{B}}$ | 2.41 | | |
| Band gap $\boldsymbol{E_g^{\Gamma-\Gamma}}$(eV) | 5.70 | | |
| Band gap $\boldsymbol{E_g^{R-R}}$(eV) | 7.70 | | |
| Band gap $\boldsymbol{E_g^{M-M}}$(eV) | 7.90 | | |
| Band gap $\boldsymbol{E_g^{X-X}}$(eV) | 7.10 | | |

I[8], II[31], III[32]